# Experimental observation of magnetoelectricity in spin ice $Dy_2Ti_2O_7$

L. Lin,[1,2] Y. L. Xie,[2] J.-J. Wen,[3] S. Dong,[1,a)] Z. B. Yan,[2] and J.-M. Liu[2,b)]

[1]*Department of Physics & Jiangsu Key Laboratory for Advanced Metallic Materials, Southeast University, Nanjing 211189, China*

[2]*Laboratory of Solid State Microstructures and Innovation Center of Advanced Microstructures, Nanjing University, Nanjing 210093, China*

[3]*Institute for Quantum Matter and Department of Physics and Astronomy, The Johns Hopkins University, Baltimore, Maryland 21218, USA*

**Abstract**

The intrinsic noncollinear spin patterns in rare-earth pyrochlore are physically interesting, hosting many emergent properties, e.g. spin ice and monopole-type excitation. Recently, the magnetic monopole excitation of spin ice systems was predicted to be magnetoelectric active, while rare experimental works have directly confirmed this scenario. In this work, we performed systematic experimental investigation on the magnetoelectricity of $Dy_2Ti_2O_7$ by probing the ferroelectricity, spin dynamics, and dielectric behaviors. Two ferroelectric transitions at $T_{c1}$=25 K and $T_{c2}$=13 K have been observed. Remarkable magnetoelectric coupling is identified below the lower transition temperature, with a significant suppression of the electric polarization upon applied magnetic field. It is surprised that the lower ferroelectric transition temperature just coincides with the Ising-spin paramagnetic transition point, below which the quasi-particle-like monopoles are populated, indicating implicit correlation between electric dipoles and spin moments. The possible magnetoelectric mechanisms have also been discussed although a decent theory remains unavailable up to date. Our results will stimulate more investigations to explore multiferroicity in these spin ice systems and other frustrated magnets.

# 1. Introduction

Spin frustration, which arises from the competitive interactions, is common in correlated electronic systems with multifold degrees of freedom [1]. Perhaps the most exciting discovery in spin frustrated systems in the past decade is the emergent multiferroicity with strong coupling between ferroelectric and magnetic orders [2-13]. Spin ice pyrochlores, such as $Dy_2Ti_2O_7$, are typical highly frustrated magnets, whose structure can be described as a corner-sharing tetrahedron network (see Fig.1(a)) [14]. Forced by the strong magnetocrystalline anisotropy, each spin points towards or away from the center of tetrahedron, i.e. Ising-like spins. Although the combination of exchange and dipolar interactions gives rise to the "2-in-2-out" configuration all over the tetrahedron network at low temperatures ($T$'s), the macroscopic degeneracy is unavoidable, resulting in an unusual finite zero-point entropy [15-19]. Furthermore, recent works on thin-film spin ices showed significant modification of the residual entropy, offering new opportunities for manipulation of the exotic magnetic properties of spin ices [20].

In addition to the zero-point entropy, another interesting magnetic issue in spin ices is the magnetic monopole excitation [21-27]. As shown in Fig.1(b), an isolated spin-flip generates a "3-in-1-out" and "1-in-3-out" topological defect pair, which breaks the magnetic charge neutrality of tetrahedron [27]. Such a defect pair, nominally magnetic monopole-antimonopole, can separate and move independently without dissipating more energy. The virtual connection between the separated monopole and antimonopole is a Dirac string [24]. Although the experimental evidence of magnetic monopoles can be traced by neutron scattering technique [24, 26], the transport of magnetic charges has been rarely investigated, probably due to the difficulty to find the appropriate macroscopic parameter to characterize magnetic charges [22-23, 25].

Very recently, Khomskii proposed a novel mechanism that each monopole (magnetic charge) carries a finite electric dipole in spin ice systems via the exchange striction mechanism [27]. In addition, another mechanism based on the Dzyaloshinskii-Moriya (DM) interaction was once proposed to generate a finite polarization for the noncollinear spin patterns [28-29]. Experimentally, our group once reported the polarization of polycrystalline Cr-doped $Ho_2Ti_2O_7$ at a relatively high $T$ (e.g. up to 140 K), while the microscopic mechanism is unknown [30-31]. Later, one more ferroelectric transition at low $T$ (~28 K) was identified in the single crystalline $Ho_2Ti_2O_7$ [32]. Saito *et al* and Katsufuji also found significant magnetodielectric effects in $Dy_2Ti_2O_7$ and $Ho_2Ti_2O_7$ at low $T$'s [33-34]. Despite these isolated evidences, a systematic experimental study of magnetoelectricity for these highly frustrated magnets remains unavailable yet.

In this work, we present detailed experiments on the ferroelectricity, dielectric property, spin dynamics, and magnetoelectric coupling. Different from previous reports, two independent ferroelectric transitions have been identified at low $T$'s (below 25 K), whose magnetoelectric responses are totally different. The available microscopic mechanisms are adopted to interpret the origin of electric polarization, although inconsistencies exist more or less. Anyhow, it can be confirmed that the measured electric polarization has two-fold origins. One is from the structural distortion sustaining at relatively high temperature, and the other one might be the magnetic related, although the underlying mechanism remains an open question and needs further studies.

## 2. Details of experimental methods

Polycrystalline $Dy_2Ti_2O_7$ was prepared by the conventional solid state method. Stoichiometric mixtures of high purity $Dy_2O_3$ and $TiO_2$ were thoroughly ground, and heated in air at 1250 °C-1400 °C with several intermittent heating and grinding steps until high quality single phase was obtained. For comparison purpose, a series of doped $Dy_2Ti_2O_7$ samples were prepared in the same conditions, including $Dy_{2-x}Gd_xTi_2O_7$, $Dy_{2-x}Tb_xTi_2O_7$ and $Y_2Ti_2O_7$.

For electric measurements, each sample was first polished into a disk with 0.2 mm in thickness and then coated with Au on each side as electrodes. The crystal structure was checked by X-ray diffraction (XRD) (Bruker Corporation) equipped with Cu $K\alpha$ radiation. All the reflections are assigned to the single-phase cubic pyrochlore structure. The dielectric measurements were performed using the HP4294A impedance analyzer associated with physical properties measurement system (PPMS) (Quantum Design, Inc.). The *ac* magnetic susceptibility (real part $\chi'$ and imaginary part $\chi''$) was measured on powder sample using the PPMS system with the ACMS option (frequency $f$=10 Hz~10 kHz).

The electric polarization ($P$) was measured by the high precision pyroelectric current method. In detail, the sample was first poled under an electric field ($E_{pole}$=10 kV/cm) and cooled down from 100 K to 2 K. Then the electric field was removed, followed by a sufficiently long time short-circuiting (60 min at 2 K, implying a relaxation of the spin structure for 60 min). During this process, the background of electrical current was reduced to the minimum level (e.g. <0.5 pA), which can maintain for more than one hour. The $P$ was obtained by integrating the pyroelectric current ($I$) through warming the sample at rates of 2 K/min~6 K/min, using Keithley 6514 electrometer connected to PPMS [35]. The presented data are taken from the measurements at the rate of 2 K/min, implying that a time scale of 14 min from 2 K to 30 K. For the pyroelectric current measurement under a constant magnetic field ($H$), the assigned field $H$ was applied to the sample at $T$=2 K before the short-circuiting process until the end of the measurements at the rate of 2 K/min. We also measured the isothermal polarization $P$ against $H$. The sample was first electrically poled and cooled down to an assigned $T$ with $H$=0, and then the electric field was removed, followed by sufficiently long time short-circuiting (60 min at 2 K, implying a relaxation of the spin structure for 60 min). The released magnetoelectric current upon increasing $H$ from $H$=0 to $H$=±9 T at a rate of 100 Oe/s were collected.

## 3. Results

### 3.1. Dielectric and ferroelectric properties

Our first concern goes to the dielectric and ferroelectric data on $Dy_2Ti_2O_7$. The dielectric constant ($\varepsilon$) at low $T$ is presented in Fig.2(a), while its insert shows the dielectric data in a broader $T$ region. The $\varepsilon(T)$ increases gradually with decreasing $T$, and interestingly, two broad anomalies are identified roughly at $T_{c1}$~25 K and $T_{c2}$~13 K respectively, implying the possible ferroelectric transitions at these points.

To confirm these ferroelectric transitions, the pyroelectric current $I(T)$ data at three different warming rates (2, 4, and 6 K/min) are plotted in Fig.2(b). Each curve shows two clear peaks right below $T_{c1}$ and $T_{c2}$, without significant peak shift (only 0.7 K shift for the high-$T$ peak between 2 K/min and 6 K/min) along the $T$-axis for different warming rates. It is noted that a tiny shift (e.g. < 1.0 K) of the $I$-$T$ curve to the high-$T$ side is inevitable due to some practical reasons. For example, during the warming process, especially for the high rate one (e.g. 6 K/min), one can't exclude the tiny temperature difference between the samples and the thermocouples attached aside the samples. The temperature difference between inside sample and thermocouples would become bigger when the warming rate is higher. Considering that the temperature step is ~ 0.1 K and the practical reasons described above, the extrinsic contribution, if any, should be very weak, and will not affect our physical conclusion. The polarizations integrated from these pyrocurrents are almost identical for different warming rates. These characters imply that the measured $I(T)$ does come from the pyroelectric effect associated with the electric polarization [36]. The complete reversal of $P$ is demonstrated by using the negative poling fields ($E_{pole}$=-10 kV/cm), as shown in Fig.2(c). The first ferroelectric transition occurs at $T_{c1}$ with a saturated pyroelectric $P$~1.5 μC/m$^2$ and the second transition appears at $T_{c2}$ with a saturated $P$~3.65 μC/m$^2$ at 2 K.

Given the fact that the Dy$_2$Ti$_2$O$_7$ exhibits two ferroelectric transitions at $T_{c1}$ and $T_{c2}$, in the following, the characteristics of these transitions will be studied by a series of specially designed experiments, including measuring the magnetoelectric response, modulating the spin-ice configurations through Dy's site substitutions, characterizing the magnetic relaxation by *ac* magnetic susceptibility.

### 3.2. Magnetoelectric response

The response of pyroelectric current $I$ to magnetic field $H$ is an important evidence of magnetoelectricity. Fig.3(a) shows the measured $I(T)$ data under $H$=0, 2, 5, 7, 9 T, respectively. A prominent feature is that the peak just below $T_{c2}$ gradually shifts towards the low-$T$ side and the peak height is suppressed with increasing $H$, while the peak just below $T_{c1}$ remains stationary even at $H$=9 T. This suggests that the ferroelectric transition at $T_{c2}$ should be magnetic relevant, while that at $T_{c1}$ may not. A possible origin for the ferroelectric transition at $T_{c1}$ is pure structural relevant rather than spin relevant.

Regarding the magnetoelectric response, the evaluated $P(T)$ curves under different $H$ are plotted in Fig.3(b). The $P$ is significantly suppressed by $H$ below $T_{c2}$, but this suppression is much weak above it. The decrease of $P$ under $H$=9 T is maximal (~25%) at $T$=2 K. In addition, we investigate the magnetoelectric response in the isothermal mode. In Fig.4(a-b), the measured $P$(time) and $P$ ($H$) at $T$=2 K is plotted, while the whole $P$ response to $H$ in one $H$ cycle between -9 T and 9 T are plotted in Fig.4(c). It is clear that during the whole cycle, $P$ is gradually suppressed upon increasing $H$, and shows remarkable path-dependent behavior. The irreversible variation of $P$ upon the $H$-cycling may be due to a sequence associated with the magnetic domain wall motion driven by varying $H$ [37].

This irreversibility of magnetoelectric response is rather weak above $T_{c2}$, evidenced by two sets of $I(T)$ data (before and after the $H$-cycling at 2 K) for the same sample, as shown in Fig.4(d). It is clear that the peak just below $T_{c1}$ remains identical for these two cases, and that just below $T_{c2}$ is partially suppressed by the $H$-cycling. One then can conclude that the ferroelectric transition at $T_{c1}$ is relevant with the magnetic structure but that at $T_{c2}$ is not.

*3.3. Roles of $Dy^{3+}$ ions*

To understand aforementioned magnetoelectricity, the role of $Dy^{3+}$ ions is the key factor. In this subsection, we will try to modulate $Dy_2Ti_2O_7$'s properties by substituting $Dy^{3+}$ ion with other magnetic or nonmagnetic rare-earth species. First, we take the nonmagnetic $Y_2Ti_2O_7$ for comparison, noting that $Y^{3+}$ ion (0.900Å in ionic radius) is only slightly smaller than $Dy^{3+}$ ion (0.912 Å in ionic radius). The measured $I(T)$ data for $Y_2Ti_2O_7$ using the same protocol are plotted in Fig.5(a) comparing with the data of $Dy_2Ti_2O_7$. Clearly, there is no any pyroelectric current detected for $Y_2Ti_2O_7$ over the whole measured $T$-range, suggesting that $Dy^{3+}$ ion is the core ingredient for the polarization generations at $T_{c1}$ and $T_{c2}$.

Second, we turn to the Tb-substituted $Dy_2Ti_2O_7$, $Dy_{2-x}Tb_xTi_2O_7$. Taking the $x$=0.5 sample for example, the pyroelectric data are also shown in Fig.5(a) for comparison. It is known that the magnetic ground state of $Tb_2Ti_2O_7$ remains a cooperative paramagnet or spin liquid state till extremely low $T$ (e.g. 0.07 K), with neither long-range Néel order nor spin glass ordering [38], substantially different from $Dy_2Ti_2O_7$'s spin ice magnetic structure. In our $Dy_{2-x}Tb_xTi_2O_7$ sample, the $Tb^{3+}$ substitution remarkably suppresses the two current peaks which disappear completely when $x$>1.0 (not shown).

Third, similar phenomena are also observed for the Gd-substituted $Dy_2Ti_2O_7$, $Dy_{2-x}Gd_xTi_2O_7$. Noting that $Gd_2Ti_2O_7$ is a typical Heisenberg spin system, since $Gd^{3+}$ ($S$=7/2) owns half-filled 4$f$ shell and zero orbital momentum, i.e. the $Gd^{3+}$ spins can rotate in three-dimensional space freely. Fig.5(b) shows the measured $I(T)$ data for several $Dy_{2-x}Gd_xTi_2O_7$ samples [39]. The two peaks are seriously suppressed and nearly disappear when $x>1.4$.

The gradual disappearance of the transition at $T_{c2}$ and $T_{c1}$ is under the expectation since the spin ice structure should be corroded by the $Tb^{3+}$ or $Gd^{3+}$ substitution. As stated before, the transition at $T_{c1}$ is possibly attributed to lattice distortion rather than the magnetic structure. Since the $Re_2Ti_2O_7$ (Re: rare earth ion) family belongs to the same space group $Fd$-$3m$, and the $Re^{3+}$ and $Ti^{4+}$ ions occupy the $16d$ and $16c$ positions respectively, the Raman active modes involve only the O ion vibrations. In fact, Raman spectroscopy on $Dy_2Ti_2O_7$ in the low $T$ range revealed a clear new mode characterizing a phonon softening below ~110 K, accompanied with a structural transformation [40-41]. No such phonon mode softening has been identified in $Gd_2Ti_2O_7$ and $Tb_2Ti_2O_7$ [40]. Therefore, these previous Raman studies also provided a hint to understand the transition at $T_{c1}$, although no firm conclusion has been made so far.

*3.4. ac magnetic susceptibility*

Now we turn to the spin-relevant ferroelectric transition at $T_{c2}$. The above data and discussions have demonstrated the correlation of this transition with the particular spin structure of $Dy_2Ti_2O_7$ in the low $T$ range. However, the microscopic details of this correlation remain an open issue. Recently, transverse field muon spin rotation (μSR) [22] and *ac* susceptibility methods [23,42-47] have been adopted to probe the dynamics of magnetic monopoles in spin ice. Here, to further enhance our understanding of the emergence of multiferroicity in $Dy_2Ti_2O_7$, a series of measurements on the *ac* magnetic susceptibility of $Dy_2Ti_2O_7$ and Gd-substituted $Dy_2Ti_2O_7$ systems are carefully performed.

The measured characteristic relaxation time ($\tau$) of $Dy_2Ti_2O_7$ was evaluated by the reciprocal of the maximum $\tau=1/f_{max}$ in the dielectric imaginary part $\chi''(f)$ curve, where $f_{max}$ is the peak frequency. The $\tau$-$T$ dependence curve is presented in Fig.6. A clear plateau region between 4 K and 13 K is identified, in which massive magnetic monopoles excite via quantum tunneling [23]. Generally, there are two contributions to magnetic monopole, one is the "single-charged" monopoles (connected through Dirac string), and the other is "double-charged" (magnetic monopole pair). At high $T$'s, the proliferation of bound defects will both disrupt existing strings and reduce the mean free path for diffusing monopoles [26]. Theoretical works [23, 42] also provides strong evidence that the "quantum tunneling" regime in the magnetic relaxation measurement can be interpreted entirely in terms of the diffusion motion of monopoles, constrained by a network of "Dirac strings".

With further decreasing $T$ below 4 K, a gradual spin freezing of 2-in-2-out configuration starts to emerge in, and fully evolve into spin ice state below 1.1 K, noting that there are massive monopole excitations in the temperature region 2 K<$T$ <4 K. Such a signature of sharp increase of $\tau$ suggests that thermal relaxation process becomes important due to the strongly correlated spin-spin interactions [44, 47]. For the high $T$ regime (>13 K), a sharp drop of $\tau$ with increasing $T$ occurs, indicating another thermal relaxation process [23].

Subsequently, for comparison, a series of $Dy_{2-x}Gd_xTi_2O_7$ samples with $x \in [0,2]$ are studied to reveal the spin dynamics in Gd-doped $Dy_2Ti_2O_7$. The aforementioned ferroelectric property shows that a slight Gd doping level $x$ can significantly suppress ferroelectric phase. Taking the $x$=1.1 sample as an example, and the real part $\chi'(T)$ (Fig.7a'-d') and $\chi''(T)$ (Fig.7a''-d'') under several magnetic fields are plotted. There is no sign for any spin freezing behavior observed within selected frequencies at zero magnetic field. However, the imaginary part of susceptibility shows a sharp increase at low temperature, which indicates a strongly dissipative process occurring below 2 K [48]. When the magnetic field increases to 1 Tesla, two clear dips at 3 K, 13 K emerge, identified as $T_f$, $T^*$, respectively. At $H$=2 T, another high-temperature peak (indicated by the $T_s$) shows up at ~ 25K. Here, the two peaks $T_f$, and $T_s$ should have the same origin as those in pure $Dy_2Ti_2O_7$, and the newly developed $T^*$ should be originated from $Gd^{3+}$, which alters the local Dy-Dy interactions and corresponding crystal field effects. Similar features are also observed in other substituted samples.

The contour plot of $\chi''(T)$ at $H$=2 T is shown in Fig.8 as a function of substitution and temperature. It can be clearly seen that $T^*$ exists in a large region with increasing $x$, while $T_s$ gradually disappeared when $x$>1.1, noting that the pyroelectric current nearly disappears when $x$>1.4. The nearly one-to-one correspondence between the appearance of polarization and $T_s$ indicates some hidden correlations between spin dynamics and ferroelectricity in spin ice.

## 4. Discussions

Although no well-defined clue on the transition at $T_{c1}$ is obtained at this stage, above experimental data demonstrate the correlation of the ferroelectric transition at $T_{c2}$ with the particular spin structure of $Dy_2Ti_2O_7$. Based on above results, one is allowed to discuss the possible origins of ferroelectric polarization below $T_{c2}$, while the origin for that at $T_{c1}$ is left for the future investigations.

*4.1. Magnetic monopole*

The magnetic monopoles and their cooperative behavior may be one of the origins for the observed magnetoelectric effects. According to Khomskii's theory, each monopole (magnetic charge) carries a finite electric dipole in spin-ice systems [27]. The electric charge transfer (i.e. electric dipole) on any site *i* can be written as:

$$\delta q_0 \propto 2\vec{S_1} \cdot (\vec{S_2} + \vec{S_3} + \vec{S_4}) - 2(\vec{S_2} \cdot \vec{S_3} + \vec{S_2} \cdot \vec{S_4} + \vec{S_3} \cdot \vec{S_4}), \tag{1}$$

where $S_i$ (*i*=1-4) is the spin. One considers four possible spin configurations as shown in Fig.9(a). No charge transfer is available ($\delta q_0$=0) for the configurations (1) (4-in or 4-out) and (4) (2-in/2-out), while finite electric charge transfers ($\delta q_0 \neq 0$) can be generated in the configurations (2) (3-in/1-out) and (3) (1-in/3-out).

Considering the electric and magnetic nature of monopole, it can be flexibly moved by external electric/magnetic fields. Take the lattice shown in Fig.9(b) as an example, where three electric dipoles are available under zero electric field. Upon a poling electric field (*E*) applied to the lattice, these dipoles will be polarized to lower the electrostatic energy. These processes can be achieved by one-step motions of the two monopoles. As shown in Fig.9(c), two spin-flip events annihilate original dipoles 2 and 3 and create dipoles 4 and 5. When these electric dipoles in the lattice are polarized, a finite macroscopic electric polarization is obtained.

According to previous neutron studies, the spins of $Dy^{3+}$ becomes Ising-type due to the strong magnetocrystalline anisotropy below 13 K [26,49], in consistent with $T_{c2}$. Therefore, the magnetism-related ferroelectric polarization emerges just in the temperature region with massive excitation of monopoles. Intuitively, there might be some correlations between monoples and ferroelectricity. This hypothesis is consistent with the tentative examination of Gd/Tb substituted $Dy_2Ti_2O_7$ as shown in Fig.5. A reasonable modulation of ferroelectricity is expected within the scenario.

However, although the poling process before pyroelectric current measurement can polarize these monopole-carried dipoles, it is unclear why their orientations can persist after the removal of electric field. The simple dipole-dipole interactions usually prefer to depolarize the macroscopic polarization. As indicated by Fig.6, the magnetic dynamics of $Dy_2Ti_2O_7$ is quite fast above 5 K. In absence of electric or magnetic field, the magnetic relaxation should be very quick in our measurement $T$ region, which should lead to the dissipation of polarization with time rapidly.

Following this line, we have conducted additional experiments to check the possible evolution of polarization with time. Fig.10(a) shows the time dependence of pyroelectric current at 2 K. It is noted that this process can sustain for at least one hour, which is very long enough to let the spin fully relaxation. Subsequently, during the pyroelectric measurement with increasing temperature from 2 K at the rate of 2 K/min, we stopped the warming of the sample and annealed it at 5 K for several times (from 1 min till 30 min) and then continued the temperature increasing. With this protocol, we did not observe any difference of the *P-T* curves, as shown in Fig.10(b) for three cycles of measurements. In this sense, the polarized state is stable, without time-dependent relaxation in such a time scale. If our observed ferroelectricity owns to the Khomskii's monopole mechanism, there should be some unrevealed couplings which maintain the polarized state somehow.

In addition, our experiment found the magnetization is almost saturated under a high magnetic field (e.g. 9 T). In this case, all spins are nearly aligned to one direction, breaking the ice rule and monopole conditions. If the ferroelectricity is caused by monopoles, the Ising-like spin pattern should be seriously destroyed by such a strong magnetic field. However, our measurement only found a moderate modulation ~25% (Fig. 3), which cannot be understood from the monopole aspect.

*4.2. Spin-orbit couplings: DM interaction & p-f hybridization*

From another point of view, it is seen that the spin chain along each of the <111> axes is noncollinear below $T_{c2}$ and the inverse DM interaction is active due to the $Dy^{3+}$-$O^{2-}$-$Dy^{3+}$ pair with bonding angle $109.47^o$ [30]. Therefore, an electric dipole can be generated for each $Dy^{3+}$-$O^{2-}$-$Dy^{3+}$ pair due to the noncollinear spin pair [28-29], as shown schematically in Fig. 9(d). The overall polarization is the summation of all these dipoles, which depends on the particular spin patterns. For example, in Fig. 9(d), the total polarization can be canceled or polarized. In this sense, the ferroelectricity relies on the noncollinear spin texture, but not the monopole excitation.

The magnetic field can align the spins and thus suppress the polarization, which qualitatively agrees with our observation under magnetic field. However, a quantitative analysis of this magnetoelectric response is unavailable. Similar to the above monopole theory, it is also unclear why the macroscopic polarization can be stabilized after the electric poling. The dynamics of spins can quickly relax the alignment of dipoles, in opposite to our observations.

The third possible scenario is the single-site *p-d* hybridization-induced polarization proposed by Arima [50]. If this mechanism can be extended to the *p-f* electrons systems, it may generate a polarization. However, since the four nearest-neighbor $O^{2-}$-$Dy^{3+}$-$O^{2-}$ angles are all $180^{°}$, the *p-f* hybridization mechanism cannot attribute to a net polarization due to the inter-cancellation of these dipoles.

In summary, although we have applied several established theories and scenarios to understand our experimental results, none of them can be fully consistent. In the current stage, very few knowledge regarding the ferroelectricity for these systems is available, which deserves more experimental and theoretical works to clarify the real mechanism.

## 5. Conclusion

In summary, we have performed systematic experimental investigation on the ferroelectricity and magnetoelectric coupling effect in spin ice $Dy_2Ti_2O_7$. Two ferroelectric transitions have been observed. One is from the structural distortion sustaining at relatively high temperature, and the other one might be the magnetic related. The origin of ferroelectricity has been discussed based on the magnetic monopole scenario and other possible mechanisms. However, the real physical mechanism remains an open question and needs more studies. Our results thus will stimulate more investigations on magnetoelectricity in spin ice systems and other highly frustrated magnets.

**Acknowledgment**

The authors would like to thank Prof. Collin Broholm, Dr. Seyed Koohpayeh and Referee 2 for helpful discussions. Research supported by the National Natural Science Foundation of China (Grant Nos. 11234005, 11374147, 11504048, 51332006, 51322206), the Jiangsu Key Laboratory for Advanced Metallic Materials (Grant No. BM2007204), Post-doctoral Science Fund of China (Grant No. 2015M571630), and Post-doctoral Science Fund of Jiangsu Province (Grant No. 1402043B). Work at IQM was supported by the US Department of Energy, office of Basic Energy Sciences, Division of Materials Sciences and Engineering under Grant No. DE-FG02-08ER46544.

**Figures**

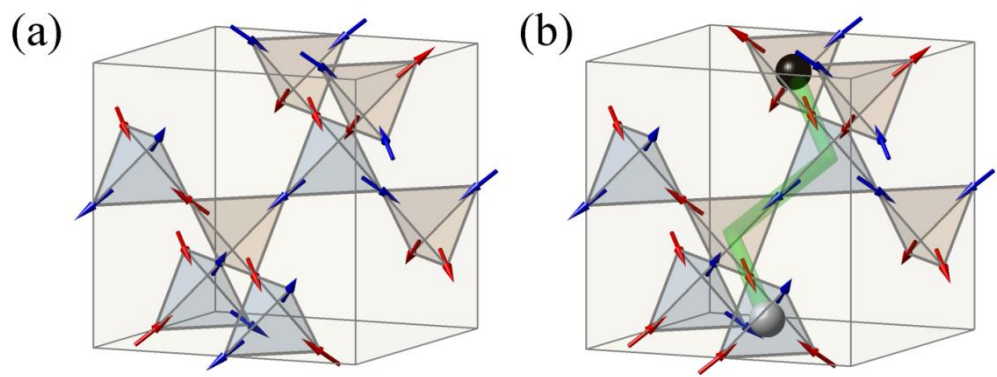

**Figure 1.** Sketch of crystalline and spin structures of pyrochlore. (a) Spin ice state with the 2-in-2-out spin configuration. (b) Magnetic monopole-antimonopole pair (black and gray spheres), linked by the so-called Dirac string (green).

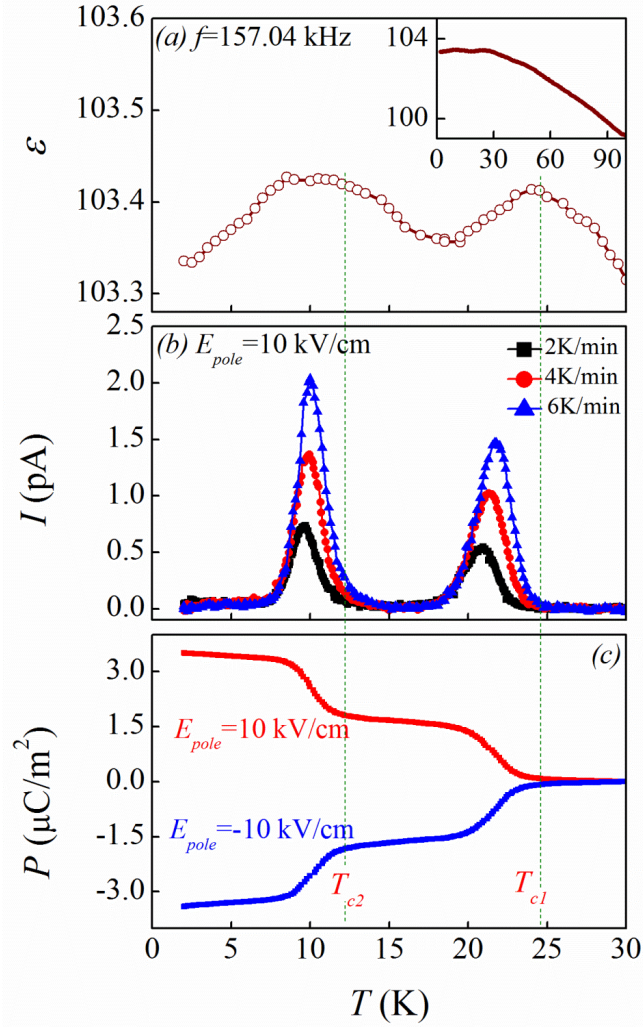

**Figure 2.** (a) The $T$-dependence of dielectric constant $\varepsilon$ from 2 K to 30 K measured at the frequency $f$=157.04 kHz. The inset shows the dielectric constant from 2 K to 100 K. The sample warming rate is 2 K/min. (b) Pyroelectric current $I$ as a function of $T$ for $Dy_2Ti_2O_7$ under a series of warming scan rates (2, 4, 6 K/min). (c) Temperature dependence of electric polarization $P$'s without magnetic field, obtained from the integration of pyroelectric currents after poling in a positive and a negative electric fields (±10 kV/cm).

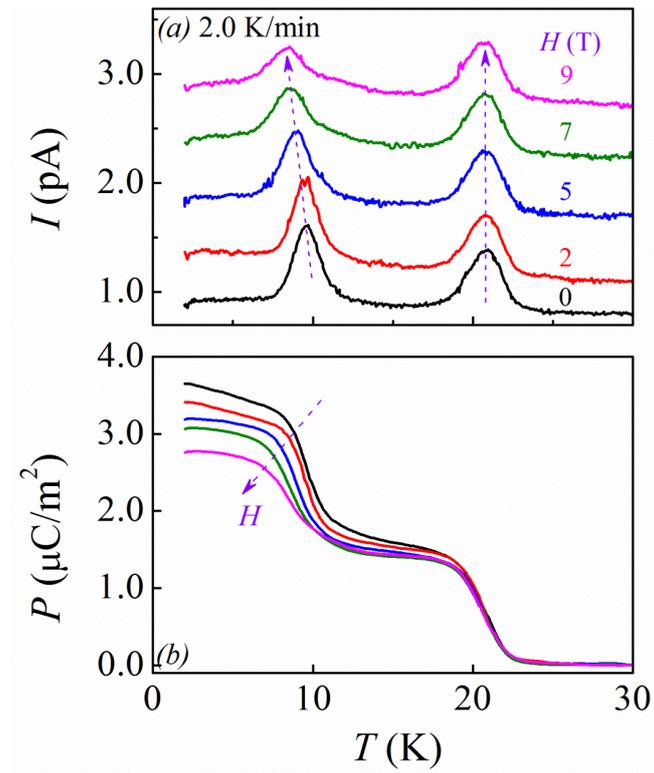

**Figure 3.** Temperature dependence of (a) pyroelectric current *I*, and (b) electric polarization *P* under several selected magnetic fields (0, 2, 5, 7, and 9 T) for $Dy_2Ti_2O_7$. All the *I-T* curves were measured at a sample warming rate of 2 K/min.

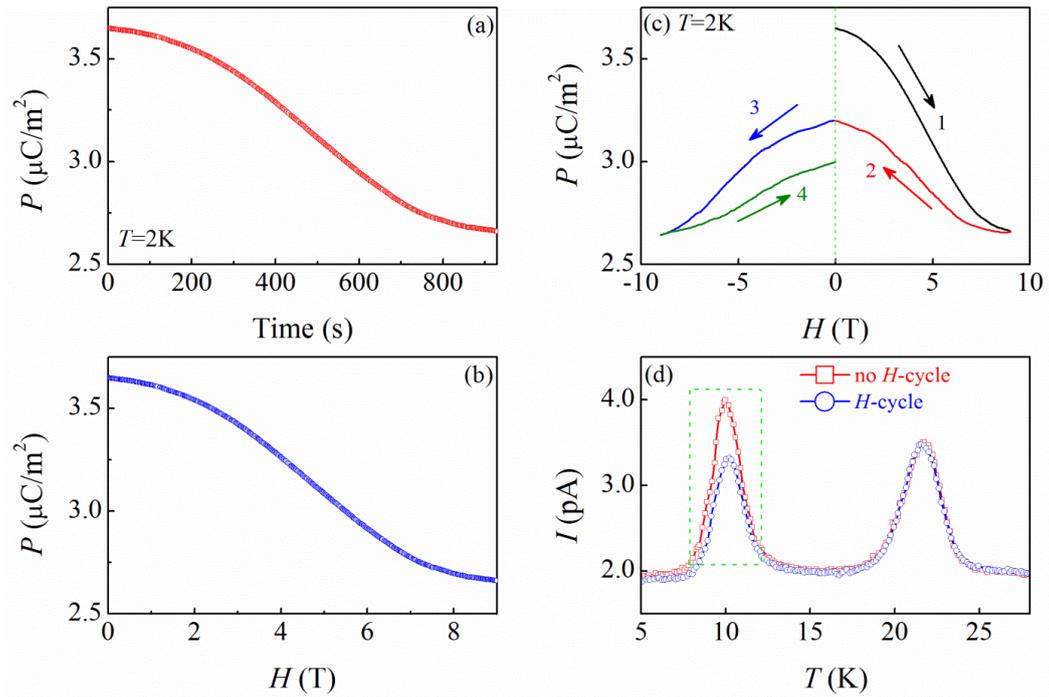

**Figure 4.** (a) Time dependence and (b) $H$ dependence of $P$ at 2 K for $Dy_2Ti_2O_7$. (c) $P$-$H$ measured under a whole $H$ cycle at 2 K. The arrows indicate the sweeping direction. (d) The red square line represents the virginal $I$-$T$ curve, while the blue square line $I$-$T$ curve is obtained by warming $T$ after an entire $H$ cycling is competed.

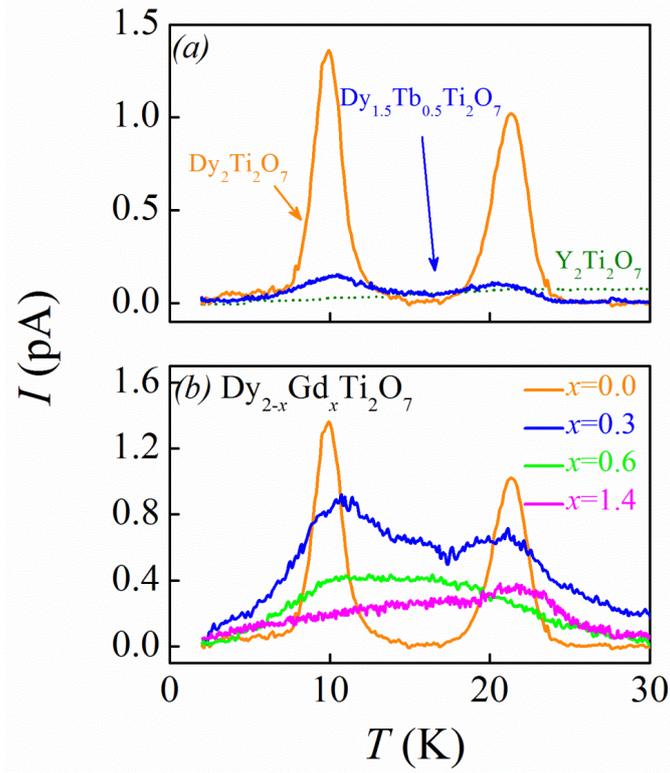

**Figure 5.** (a) Measured *I-T* curve of $Y_2Ti_2O_7$, $Dy_{1.5}Tb_{0.5}Ti_2O_7$, and $Dy_2Ti_2O_7$. (b) *T*-dependent current *I* of a series of samples $Dy_{2-x}Gd_xTi_2O_7$ (*x*=0, 0.3, 0.6, and 1.4). The sample warming rate is 4 K/min.

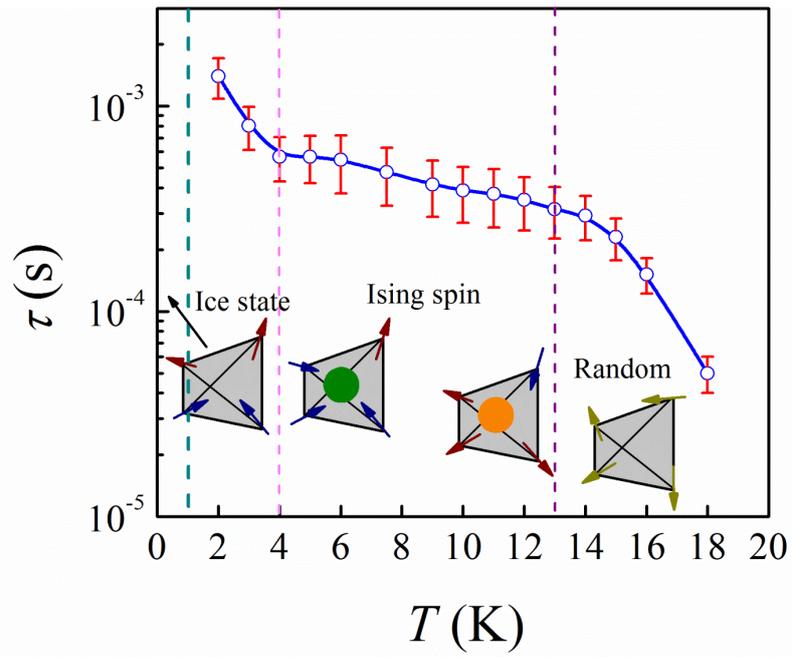

**Figure 6.** The characteristic relaxation time $\tau$, as a function of $T$ in absence of magnetic field.

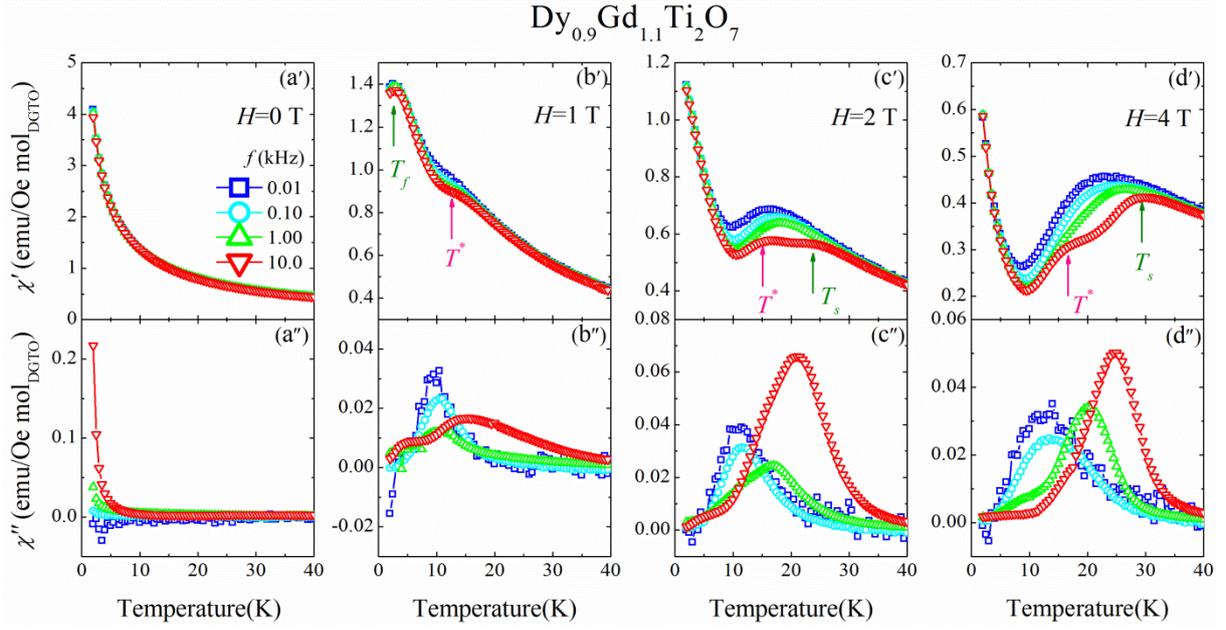

**Figure 7.** The upper (a′-d′) and lower (a″-d″) panels present the real and imaginary part of *ac* susceptibility of $Dy_{0.9}Gd_{1.1}Ti_2O_7$ at selected magnetic fields. $T_f$, $T^*$, and $T_s$ denote the low temperature spin freezing into ice state, the freezing peak associated with $Gd^{3+}$, and the single-ion peak associated with $Dy^{3+}$, respectively.

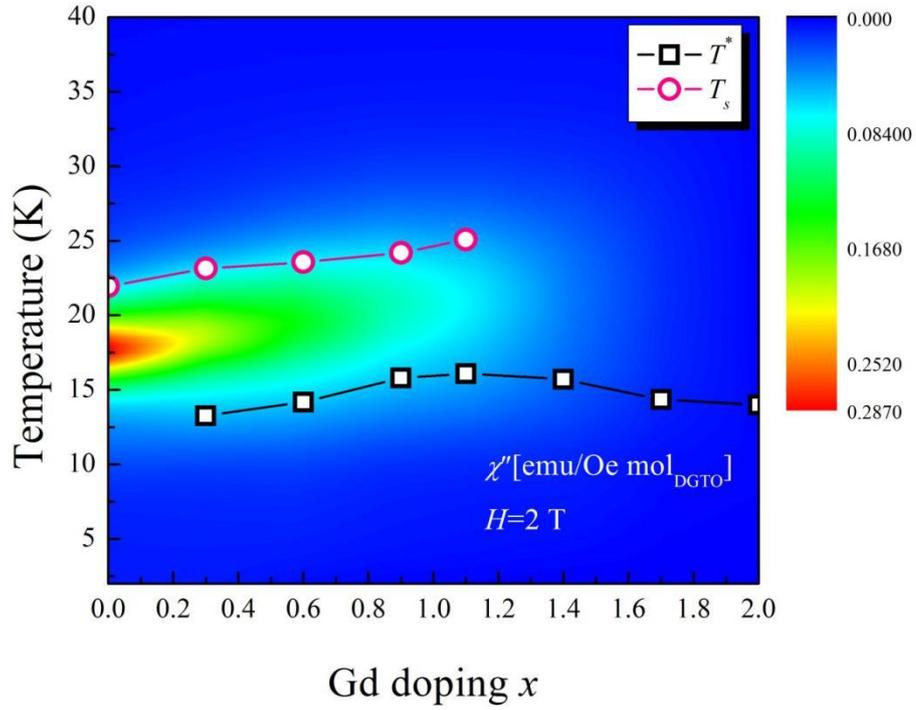

**Figure 8.** Contour plot of imaginary part of *ac* susceptibility of $Dy_{2-x}Gd_xTi_2O_7$ samples at $H$=2 T measured at $f$=10 kHz. The $x$-$T^*$ and $x$-$T_s$ relations are shown.

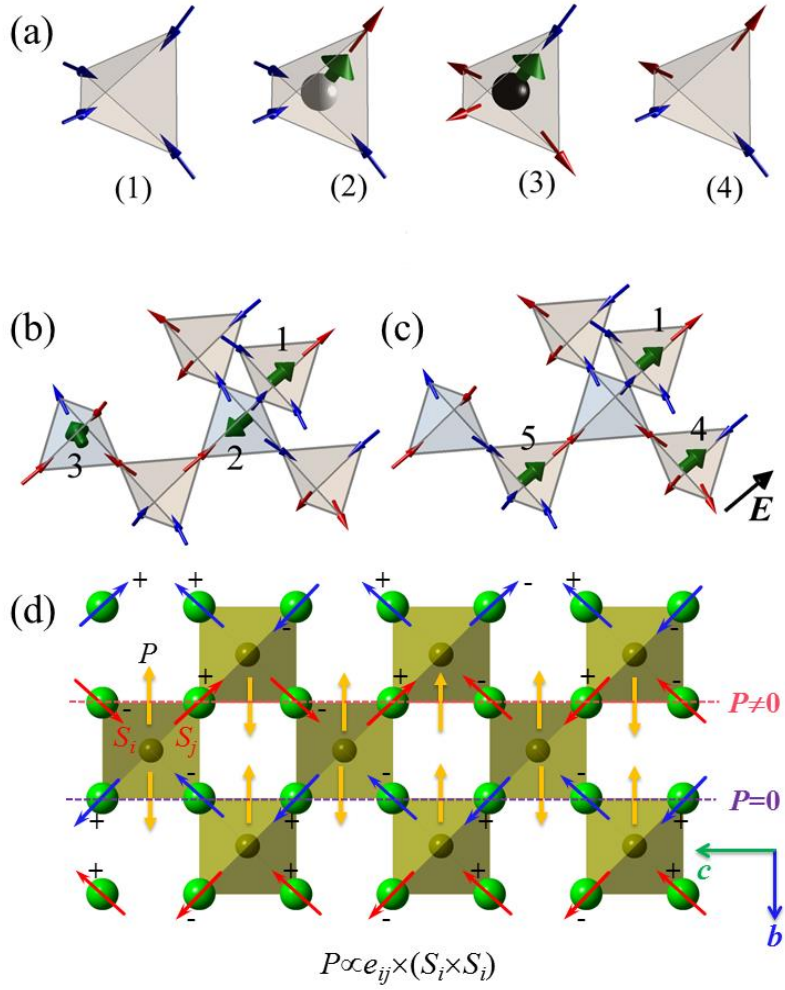

**Figure 9.** (a) The specific spin configurations of tetrahedra: (1) 4-in or 4-out, (2) 3-in-1-out, identified as magnetic monopole, (3) 1-in-3-out, as antimonopole, and (4) 2-in-2-out (the spin ice state). Only the cases (2) and (3) carry an electric dipole, indicted by the green arrows. The schematic diagram of the monopole distribution under (b) zero electric field ($E$=0), and (c) finite electric field ($E$≠0). (f) The schematic diagram of spin ice state of $Dy_2Ti_2O_7$, projected down the $a$ axis. The "+" and "-" sign indicates that the component of each spin is parallel or antiparallel to the $a$ axis. The yellow arrow presents the generated polarization between the neighbor spins $S_i$ and $S_j$ by inverse DM interaction.

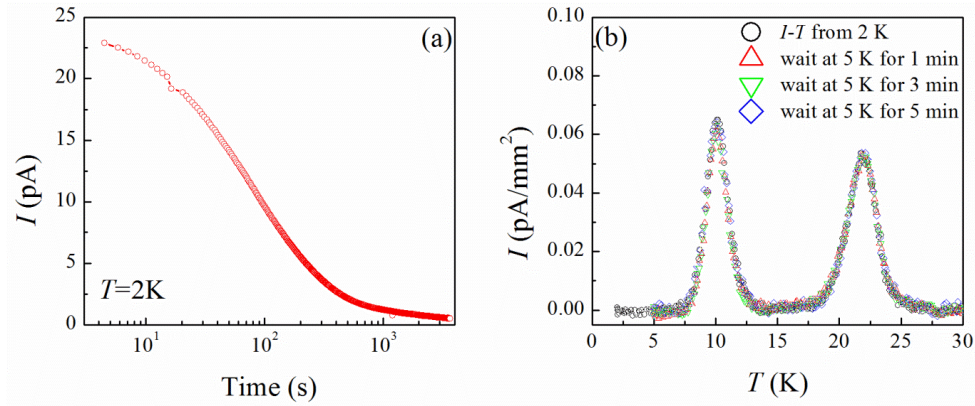

**Figure 10.** (a) The time dependence of current within the electrically short-circuiting process at 2 K. (b) The measured typical $T$-dependent pyroelectric current $I$ (black circle), and $I$-$T$ curves after relaxations at $T$=5 K for 1, 3, and 5 minutes. The sample warming rate is 2 K/min.